\documentclass[article]{agujournal2019}

\usepackage{url} 
\usepackage[inline]{trackchanges} 
\usepackage{amsmath}
\usepackage{soul}
\usepackage{caption}
\usepackage{subcaption}

\draftfalse

\journalname{Space Weather}

\begin{document}

%
%


\title{Elucidating the Grey Atmosphere: SHAP Value Analysis of a Random Forest Atmospheric Neutral Density Model}

%
%




\authors{C. Bard\affil{1}, K. Murphy\affil{2}, A. Halford\affil{1}}

\affiliation{1}{Heliophysics Science Division, NASA Goddard Space Flight Center, Greenbelt, MD}
\affiliation{2}{Independent Contractor, Thunder Bay, ON}




\correspondingauthor{Chris Bard}{christopher.bard@nasa.gov}


\begin{keypoints}
\item We apply the TreeSHAP method to provide insight for an atmospheric density Random Forest model.
\item The model recognizes that solar irradiance is a key contributor to thermospheric density variations.
\item The model is able to distinguish between day and night, dawn and dusk, and capture storm-time influence as parameterized by SYM-H.
\item The SHAP analysis suggests defining geomagnetic ``storm-time'' as periods when SYM-H falls below -60 nT.
\end{keypoints}

%
%

%
%


\begin{abstract}
We apply SHAP (SHapley Additive exPlanations) analysis using the TreeSHAP algorithm to a Random Forest model (RANDM) designed to predict thermospheric neutral density based on solar-terrestrial data. 
The analysis shows that RANDM identifies solar irradiance as a significant predictor of thermospheric density. Additionally, the model differentiates between magnetic local times, finding that dusk sectors have higher densities than dawn sectors, in line with prior research.
When comparing storm and quiet-time conditions, we find these trends persist regardless of geomagnetic activity levels. 
The analysis further demonstrates that larger geomagnetic disturbances during storms, as parameterized by the SYM-H index, are associated with higher neutral densities. 
Notably, SYM-H begins to have the overall largest contribution to density prediction among model inputs at a threshold of -60 nT. 
This suggests a quantitative definition where ``storm-time'' begins at SYM-H $< -60$ nT.
Overall, using TreeSHAP enhances our understanding of the factors influencing thermospheric density and demonstrates the value of explainable machine learning techniques in space weather research, enabling more interpretable models.

\end{abstract}

\section*{Plain Language Summary}
Statistical models based on data, such as random forests, are useful for making predictions. However, they are not as good for understanding the underlying physical system. It is not obvious which parts of the input data or model features are most important and have the largest effects on the final model prediction. Here, we use a SHapley Additive exPlanations (SHAP) analysis to look more closely at a random forest model that predicts the atmospheric density at 400 kilometers above the Earth. SHAP allows us to quantify what data the model considers most important for delivering an accurate prediction. Our analysis confirms that solar irradiance is generally the most important factor. However, the model also associates periods of increased geomagnetic activity with higher densities. Our analysis suggests that ``storm-time'' can be defined as when the geomagnetic activity, as measured by the SYM-H parameter, drops below -60 nano-Tesla. Using SHAP, we can see how the statistical model captures well-understood processes, such as the importance of solar irradiance for neutral density changes. We can also identify when geomagnetic storms may impact neutral density.

%
%

\section{Introduction}
The use of machine learning (ML) techniques within Heliophysics has recently exploded, especially for space weather applications.
However, though these models are in general successful in making predictions, they are not as useful for understanding physical relationships between model features and targets (input and predicted variables, respectively). 
Connections between the features and target data are generally hidden away in the model, which remains opaque.

Several techniques have been developed for ``explainable" ML, including SHAP (SHapley Additive exPlanations) values \cite{shapley1953,lundberg2017,lundberg2020}, Individual Condition Expectation plots \cite{goldstein2015}, and Local Interpretable Model-agnostic Explanations (LIME; \citeA{ribeiro2016}). Here, we focus on SHAP values, specifically using the TreeSHAP algorithm \cite{lundberg2020, yang2021} for random forests. Using TreeSHAP allows us not only to analyze individual predictions, but also to aggregate many individual analyses into a broad model assessment. This gives us physical insights into both local events and global behaviors. SHAP values have a heritage in Heliophysics, previously being used to investigate the radiation belts \cite{ma2023,ma2024}, ground induced currents \cite{coughlan2023}, and connections between active regions and solar flares \cite{cavus2025arxiv}. Discussions on other interpretable ML techniques can be found in, e.g., \citeA{molnar2025}.

In this paper, we apply SHAP value analysis to the Random Forest Thermospheric density model of \citeA{murphy2025} (called RANDM), demonstrating how this technique can help understand how the model works and suggest avenues for model refinement and enhancement.
By revealing feature importance and interaction patterns, SHAP values help identify which of the given input parameters most significantly influence thermospheric density predictions.
This enables an iterative development where physical understanding can be used to fill in gaps in model ``understanding" to progressively enhance model performance.
Such transparency is particularly valuable for atmospheric density forecasting, where improved models directly translate to higher fidelity orbit propagation and more accurate tracking of the growing population of satellites and debris in low-Earth orbits (LEO) (e.g., \citeA{berger2020}).
These are critical capabilities for collision prediction and avoidance in increasingly congested orbital environments.

We give a brief overview of the RANDM model, its input data features, and the TreeSHAP technique in Section \ref{sec:overview}. 
Section \ref{sec:macro} discusses insights obtained from a global model assessment, including interaction effects between features (how features may \textit{work} together to change model predictions).
Local assessments on selected single events are performed in Section \ref{sec:local}, including expansions to full-Earth predictions (variations in magnetic local time and latitude).
Finally, we recap the results in Section \ref{sec:conclude}.

\section{Data and Methods}
\label{sec:overview}
\subsection{Model and Data}
We analyze the Random forest Atmospheric Neutral Density Model (RANDM) from \citeA{murphy2025}, specifically their best-performing variant: FISM2-GEO.
RANDM is trained on solar irradiance and geomagnetic indices to predict neutral density at an altitude of 400 kilometers.
The neutral density data are derived from high-precision accelerometers aboard the dual satellites of the Gravity Recovery and Climate Experiment (GRACE) \cite{wahr2004}, as detailed in \citeA{sutton2005,sutton2009}, and normalized to 400km using NRLMSIS \cite{emmert2021nrlmsis}.
Each in situ measurement corresponds with a atmospheric location, characterized by the satellite's latitude and magnetic local time (MLT). 
These spatial parameters are also provided to RANDM using the satellite latitude and the sine and cosine of MLT.
Solar irradiances are provided by the Flare Irradiance Spectral Model 2 (FISM2) \cite{solomon2005,chamberlin2020}; the model specifically uses the 1.3nm, 43nm, 85.55nm, and 94.4nm spectral bands.
RANDM also incorporates the SYM-H and AE geomagnetic indices provided from the OMNI data \cite{king2005} supplied by NASA's Space Physics Data Facility.
Additionally, a database of geomagnetic storms between 2002 and 2012, enables us to separate the data into quiet-time, storm-time, and storm phase (main or recovery) following the methodology outlined in \cite{murphy2018,murphy2020}.

All examples in this paper are derived from the ``test" data sample of \citeA{murphy2025}, which uses the GRACE B neutral densities.
For more details on the overall dataset and the selection of the specific RANDM (FISM2-GEO) model input features, please refer to \citeA{murphy2025}.

\subsection{SHAP Values and TreeSHAP Explainer}
In this investigation we use the TreeSHAP class from the open-source SHAP python package \cite{lundberg2020} to provide explanations for RANDM. 
TreeSHAP is an algorithm designed to efficiently compute exact SHAP (SHapley Additive exPlanations) values \cite{lundberg2017} for tree-based machine learning models, such as decision trees and ensemble methods like random forests. The SHAP values are based on Shapley values \cite{shapley1953}, which measure the feature contribution to an individual model prediction relative to a baseline value.
Quantitatively, this can be expressed as \cite{lundberg2020, molnar2025}:
\begin{align}
g(x) = \phi_0 + \Sigma_{i}^N \phi_i
\end{align}
where $g(x)$ is the prediction made by model $g$ given a feature vector $x = [x_0, x_1, ..., x_N]$ with $N$ features. $\phi_0$ is the expected value of the prediction across all data vectors, while $\phi_i$ is the SHAP value of feature $i$.
For our random forest example, $\phi_0 = 1.307$.
This can be thought of as the answer to the question, ``If I don't know any feature values, what should the predicted density be?"

When provided feature values $x_i$, the model will make a prediction that is (most likely) different from the baseline value $\phi_0$. 
This raises the natural question: how did each individual feature value influence this prediction?
SHAP values are a way to measure this contribution: $\phi_i$ represents the contribution of $x_i$ to the difference between the prediction and the baseline ($g(x) - \phi_0$).
The magnitude of a SHAP value indicates the explanatory power the model assigns to that feature.
Larger values signify greater contributions to the resulting prediction, though it is possible for multiple feature SHAP values to cancel each other out.
It is important to note that SHAP values do not represent quantitative relationships in the sense that increasing an input feature by a certain amount will proportionally increase the output prediction.
Instead, SHAP values indicate how the model assesses the input features $x_i$ in the context of the input data vector $x$ when making its final density prediction.
We note that SHAP values are not used to assess model accuracy; other metrics, like R2 score or mean squared error, are better suited for that purpose.

Normally, computing SHAP values directly involves summing over all possible combinations of feature subsets, which becomes computationally infeasible for models with many features due to the exponential growth in the number of subsets. 
TreeSHAP addresses this challenge by exploiting the structure of decision trees to reduce computational complexity from exponential to polynomial time. 
Specifically, it leverages the fact that a feature's contribution can be calculated by traversing the tree paths and considering how splits on that feature affect the prediction.
The algorithm works by recursively traversing the tree, keeping track of the proportion of the training data that reaches each node and the associated feature contributions. 
For ensemble models, SHAP values are computed for each tree individually and then summed, utilizing the linearity property of Shapley values. 
TreeSHAP can also compute SHAP interaction values, which decompose the SHAP values into main effects and pairwise interaction effects between features. 
This is achieved by calculating the difference between the SHAP values when a feature pair is considered together versus when each feature is considered independently.

We note that one drawback to TreeSHAP is that, if input features are highly correlated, TreeSHAP may attribute non-zero SHAP values to all individual features even if only one should get all of the credit \cite{janzing2020, sundararajan2020}. Indeed, this occurred in the SHAP analysis here; fortunately, there are methods to check for this (see Sections \ref{sec:macro} and \ref{sec:crossInt} for further discussion). For more details on SHAP and Shapley values, see \cite{lundberg2017,lundberg2020}. Additionally, \citeA{ma2023} provide an excellent introduction to SHAP values (specifically DeepSHAP for deep neural networks) in a heliophysics application, while \citeA{molnar2025} provide a simplified explanation for TreeSHAP. Finally, because TreeSHAP can be computationally intensive for random forests, we use the Fast TreeSHAP (FTS) package \cite{yang2021}, which extends the SHAP package to enable parallel computation of individual SHAP values.

\section{Macroscale Assessment}
\label{sec:macro}
\begin{figure}
\includegraphics[width=0.9\textwidth]{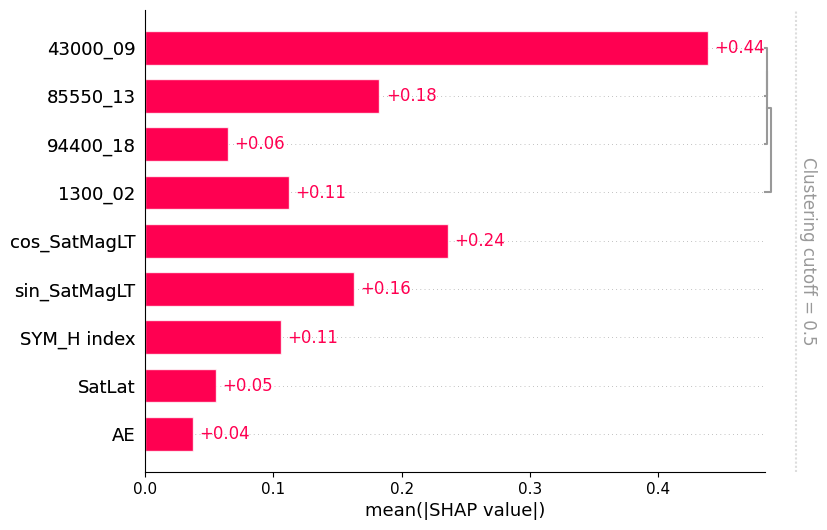}
\caption{Feature importance across 4,000 randomly selected events (2,000 storm-time and 2,000 quiet-time events), as measured by average absolute SHAP value for each feature.
The dendrogram on the right displays hierarchical clustering based on feature redundancy analysis, highlighting the potential redundancy among FISM2 spectral bands within the Random Forest model.
\label{fig:barAll}}
\end{figure}
      
\begin{figure}
\includegraphics[width=0.9\textwidth]{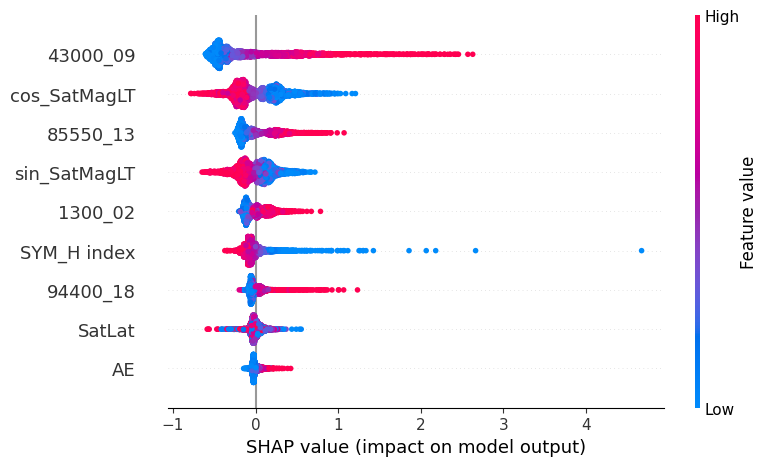}
\caption{SHAP beeswarm plot showing feature importance distribution across 4,000 randomly selected events (2,000 storm-time and 2,000 quiet-time events). 
Individual dots represent specific events, with clustering indicating the frequency of similar SHAP values. 
The horizontal position of each dot shows the impact of that feature on the model output, while color indicates the feature's value (red = high, blue = low). 
Columns are sorted in order of absolute mean SHAP value, with higher values (more important) at the top. \label{fig:beeswarmAll}}
\end{figure}

To analyze feature importance, we calculated SHAP values for each input parameter across the data.
Given the computational constraints of processing our full dataset (271,092 points), we instead analyzed representative samples of 2,000 points from each classification category: storm, non-storm, main-phase, and recovery.
Multiple random samplings confirm the consistency of our results across different 2,000-point selections.

Figures \ref{fig:barAll} and \ref{fig:beeswarmAll} show the aggregate SHAP analysis for the combined storm and quiet samples.
The SHAP analysis revealed only minor distinctions between the storm and non-storm samples, as well as between main-phase and recovery, so these are not shown here.
However, as we will show later, the SYM-H SHAP value does generally increase with storm strength.

The analysis uncovered significant overall correlations between solar-terrestrial parameters and model predictions.
Most notably, the 43 nm spectral band emerged as the dominant factor in density predictions, indicating its role as a proxy for solar energy input. 
This aligns with the findings from \citeA{murphy2025} that solar activity is likely the key factor controlling the background level of atmospheric neutral density.

Figure \ref{fig:beeswarmAll} illustrates that all FISM2 spectral bands exhibit a nearly monotonic relationship with their SHAP values: lower irradiance corresponds to lower SHAP values, while higher irradiance corresponds to higher SHAP values. 
Essentially, the RANDM model establishes an average level of solar irradiance associated with a baseline atmospheric density. 
Deviations from this average irradiance cause the predicted density to increase or decrease relative to this baseline.

Indeed, all spectral bands demonstrated positive correlations with density, though a hierarchical clustering analysis of the input features revealed partial redundancy among the FISM2 bands within the RF model.
The clustering analysis utilized univariate XGBoost tree models trained against density predictions, with each tree corresponding to a single input feature (using function \texttt{hclust} from the SHAP package).
By examining how well one feature's tree could predict the output of another's, we are able to identify redundant relationships. 
This redundancy is expected given that all spectral bands originate from the same source - the Sun, and suggests that future models may achieve similar performance while reducing model complexity by limiting the number of solar features.

Next, we find that Magnetic local time (MLT) components show substantial influence, with clear hemispheric distinctions in SHAP values. 
Using midnight as MLT = 0, positive cos(MLT) values correspond to nighttime, while negative values indicate daytime. 
The SHAP analysis demonstrated higher densities (positive SHAP) on the dayside and lower densities (negative SHAP) on the nightside, with a notable asymmetry between dawn and dusk regions. 
This pattern reflects the direct solar irradiance on Earth's dayside, where energy absorption occurs.
The dawn-dusk asymmetry is consistent with prior studies \cite{kwak2009, grocott2012, forster2017} that associate higher (lower) dusk (dawn) densities with convection cells driven by $\vec{E}\times\vec{B}$ ion drifts in the ionosphere.

These relationships between most solar-terrestrial parameters and density predictions remain consistent across the full dataset, regardless of geomagnetic conditions (storm-time vs. quiet time) or storm phases (main-phase vs. recovery). 
The only notable distinction between storm and quiet-time conditions appears in the SYM-H SHAP values, which show a higher mean and relative magnitude in the storm time dataset.
This pattern is illustrated in Fig. \ref{fig:beeswarmAll}, where the most negative magnitudes of SYM-H correspond to the highest SHAP values in the overall distribution.
These findings support the interpretation that solar irradiance universally establishes the baseline thermospheric density, while magnetospheric drivers exert their primary influence during storm conditions.

\begin{figure}
      \begin{subfigure}[b]{0.48\textwidth}
            \centering
            \includegraphics[width=\textwidth]{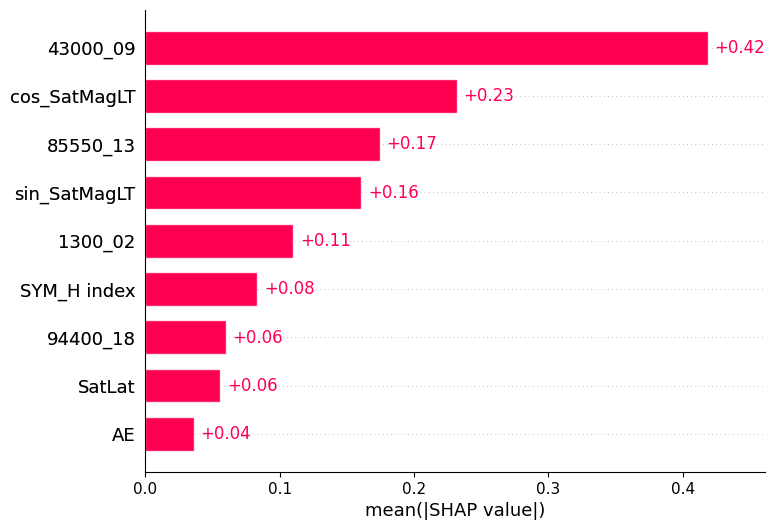}
            \caption{Small Storm: SYM-H$>-50$nT}
            \label{fig:SML_SYMH_feat}
      \end{subfigure}
      \hfill
      \begin{subfigure}[b]{0.48\textwidth}
            \centering
            \includegraphics[width=\textwidth]{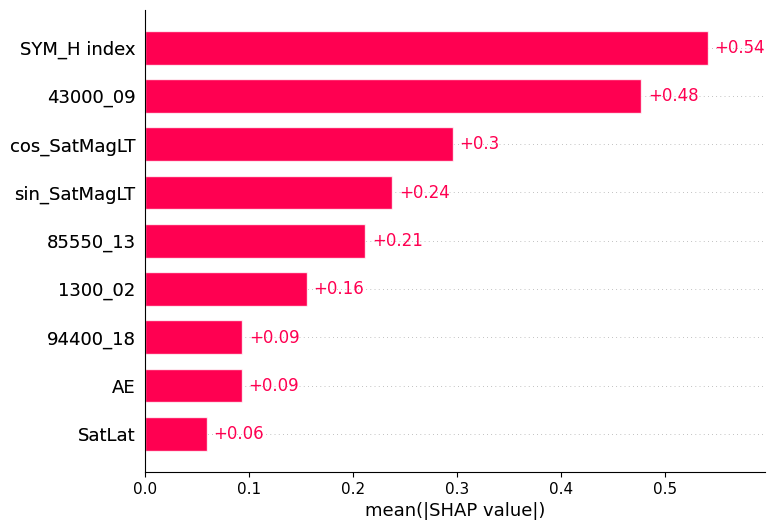}
            \caption{Moderate Storm: $-50>$SYM-H$>-100$nT}
            \label{fig:MOD_SYMH_feat}
      \end{subfigure}
      \newline
      \centering
      \begin{subfigure}[b]{0.48\textwidth}
            \centering
            \includegraphics[width=\textwidth]{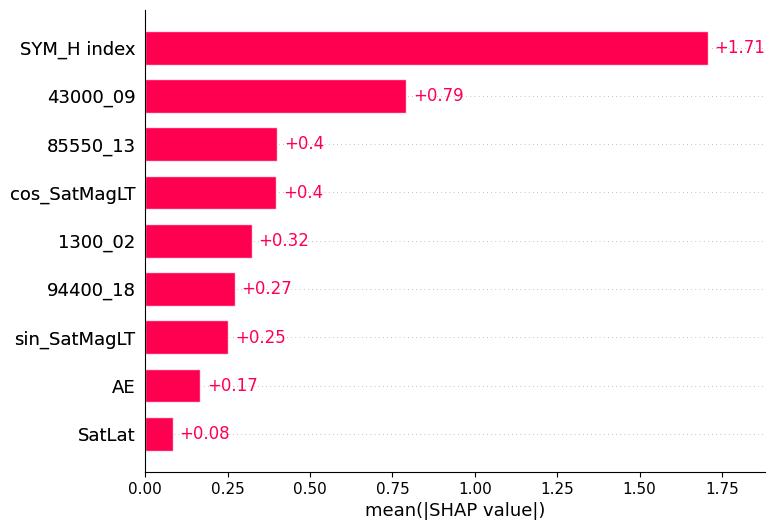}
            \caption{Large Storm: SYM-H$<-100$nT}
            \label{fig:LG_SYMH_feat}
        \end{subfigure}
        \caption{Calculated Feature importance for small, moderate, and large geomagnetic storms. The relative importance of SYM-H, as assessed by SHAP, increases as the storm gets stronger.
        \label{fig:SYMH_feat}}
\end{figure}

We can quantify this distinction further by checking the feature importance for different storm strengths, as defined by SYM-H.
Following \citeA{murphy2025}, we filter the dataset into ``small`` ($0>$ SYM-H $>-50$nT), ``moderate'' ($-50>$ SYM-H $>-100$nT), and ``large'' (SYM-H $<-100$nT) conditions, taking a sample of 2000 points from the small and moderate storm datasets, and analyzing the full large-storm dataset (1167 points).
As demonstrated in Figure \ref{fig:SYMH_feat}, it is clear that SYM-H has an increasing influence on the model forecast as storms get stronger.
We note that the SHAP values for the small storm sample (Fig. \ref{fig:SML_SYMH_feat}) are nearly identical to those calculated for the quiet time sample above.

This assessment provides insights for defining geomagnetic storm thresholds using SYM-H values.
Previously, \citeA{gonzalez1994} established storm thresholds at Dst indices of $-30$, $-50$, and $-100$ nT, corresponding to weak, moderate, and intense storms, respectively. More recently, \citeA{hutchinson2011} defined a more conservative storm threshold of SYM-H $<-80$ nT. This stricter criterion was designed to eliminate potential misidentification of weak storms with other geophysical processes that produce minor ring current enhancements, such as reconnection events unrelated to storms and substorms.

\begin{figure}
      \includegraphics[width=0.9\textwidth]{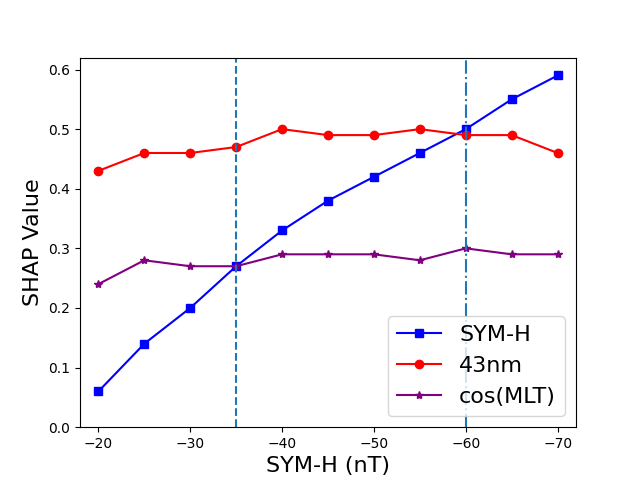}
      \caption{Average SHAP values for SYM-H (blue), 43nm FISM2 band (red), and cos(MLT) (purple) are plotted across 5 nT bins of SYM-H values from -20 nT to -75 nT. The crossover point where magnetospheric drivers (SYM-H) become more influential than solar drivers (43nm) occurs at around -60 nT (right dot-dashed line). A proposed ``storm warning'' threshold (left dashed line) is identified at approximately -35 nT, where the SYM-H SHAP value reaches greater than half the magnitude of the 43nm SHAP value, indicating significantly enhanced magnetospheric influence compared to quiet-time conditions.
      \label{fig:binned_SHAP_threshold}}
\end{figure}

Following our SHAP analysis, we suggest that ``storm time'' can be characterized as conditions where magnetospheric drivers become more important than solar drivers in determining atmospheric density. To quantify this transition point, we sorted data into 5 nT bins ranging from [-20, -25) to [-70, -75) nT, sampling 500 data points from each range. As shown in Figure \ref{fig:binned_SHAP_threshold}, the crossover point where SYM-H SHAP values exceed those of the 43nm band occurs between the [-55, -60) and [-60, -65) nT bins. This storm threshold of approximately -60 nT is consistent with, though slightly higher than, the moderate threshold of -50 nT suggested by \citeA{gonzalez1994}.

Extending this approach, we can establish a hierarchical threshold system for geomagnetic storms, akin to terrestrial tornado alerts. As above, we define a ``storm emergency'' at SYM-H $<-60$ nT. Next, we define a ``storm warning'' threshold at SYM-H $<-35$ nT: the point at which magnetospheric influence begins to significantly exceed quiet-time levels. During quiet conditions, the average SYM-H SHAP value is 0.08, about one-fifth of the 43nm SHAP value (0.42). The warning threshold occurs when the SYM-H SHAP value reaches half the 43nm SHAP value, which happens in the [-35, -40) nT bin. Here, the SYM-H SHAP value (0.27) exceeds half of the 43nm SHAP value (0.47). Notably, this is also the bin where SYM-H overtakes cos(MLT) as the second most important feature. Finally, a ``storm watch'' threshold can be defined at SYM-H $< -20$ nT: the point where SYM-H SHAP values begin to exceed quiet-time levels (SHAP $\geq$ 0.08).

Finally, our feature importance analysis using SHAP values aligns with \citeA{murphy2025}'s findings, which are based on assessing the mean decrease in accuracy when shuffling feature values. 
Both studies indicate that the FISM2 43nm band and the cos(MLT) parameter are the most influential in the RANDM model, while Satellite Latitude, FISM2 94.40nm, and AE have the least impact.

\subsection{Interaction Effects between Features}
\label{sec:crossInt}
\begin{figure}
      \includegraphics[width=0.9\textwidth]{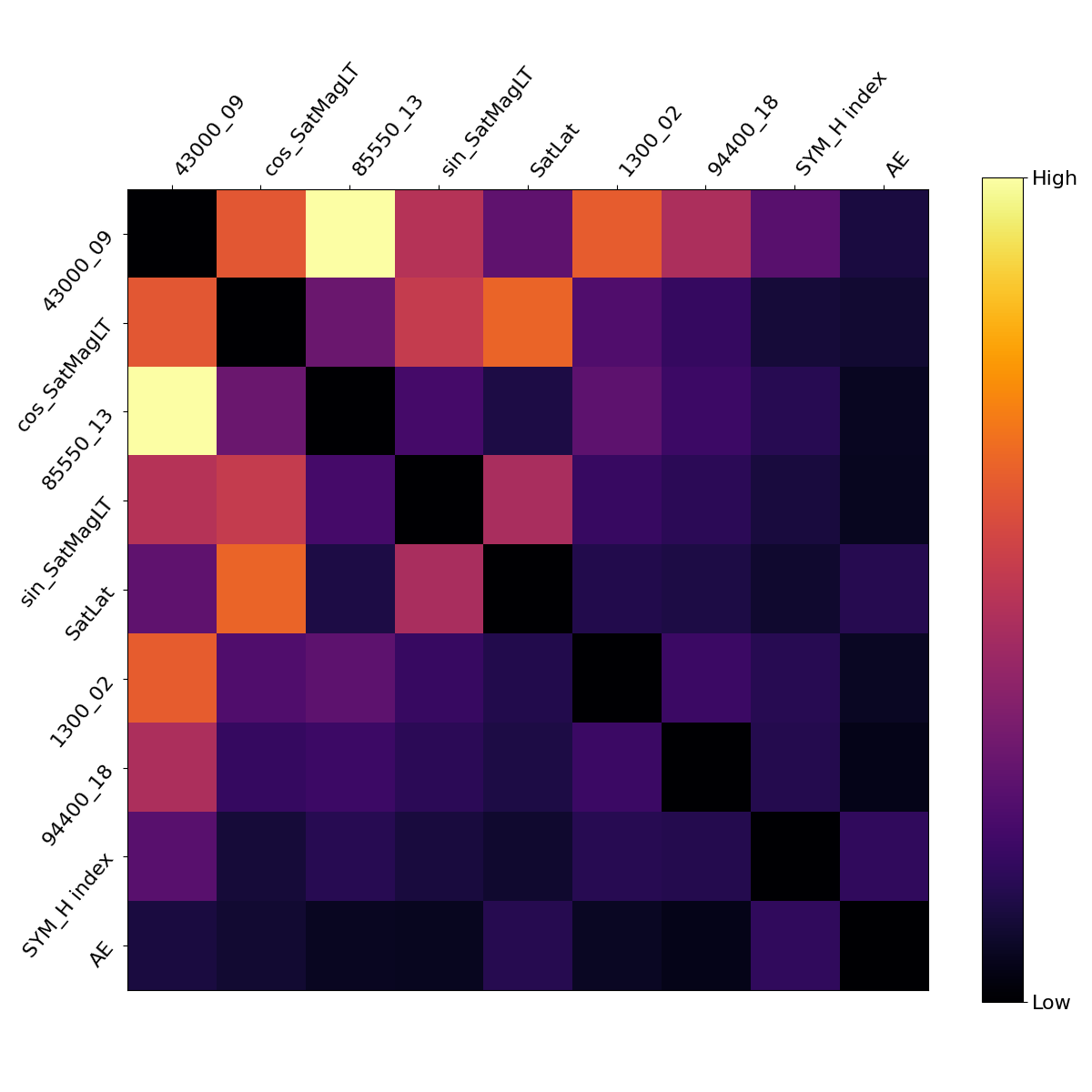}
      \caption{Interaction matrix obtained via summing the absolute values of the cross-interaction SHAP for each of the 4000 events in the combined storm+quiet data sample.
      Lighter colors indicate a high cross-interaction; dark colors indicate little cross-interaction (or that the cross-interaction is not relatively important).
      \label{fig:interaction}}
\end{figure}

\begin{figure}
      \begin{subfigure}[b]{0.48\textwidth}
            \centering
            \includegraphics[width=\textwidth]{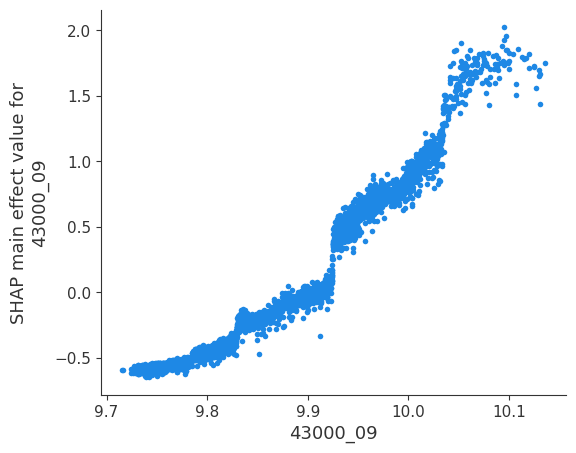}
            \caption{Main Feature for 43 nm}
            \label{fig:43main}
      \end{subfigure}
      \hfill
      \begin{subfigure}[b]{0.48\textwidth}
            \centering
            \includegraphics[width=\textwidth]{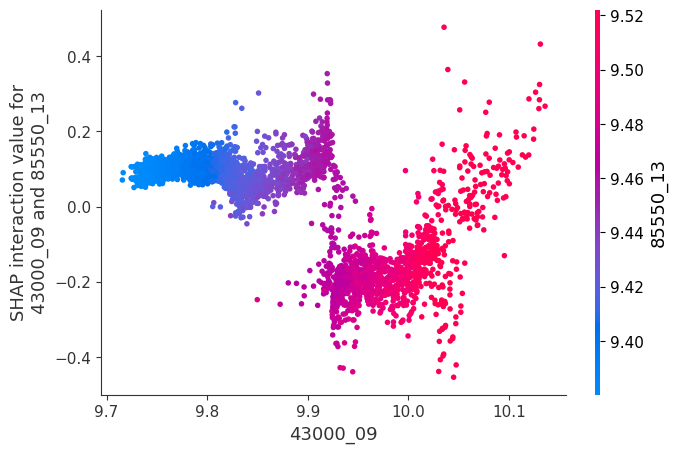}
            \caption{43nm vs. 85.55nm}
            \label{fig:4385cross}
      \end{subfigure}
      \newline
      \begin{subfigure}[b]{0.48\textwidth}
            \centering
            \includegraphics[width=\textwidth]{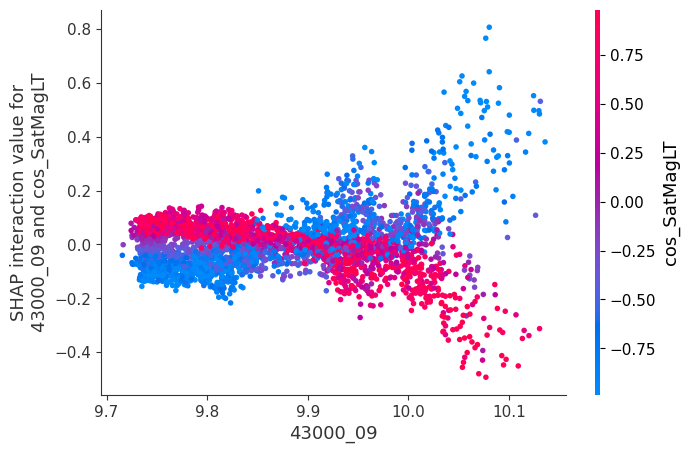}
            \caption{43nm vs. cos(MLT)}
            \label{fig:43coscross}
        \end{subfigure}
        \begin{subfigure}[b]{0.48\textwidth}
            \centering
            \includegraphics[width=\textwidth]{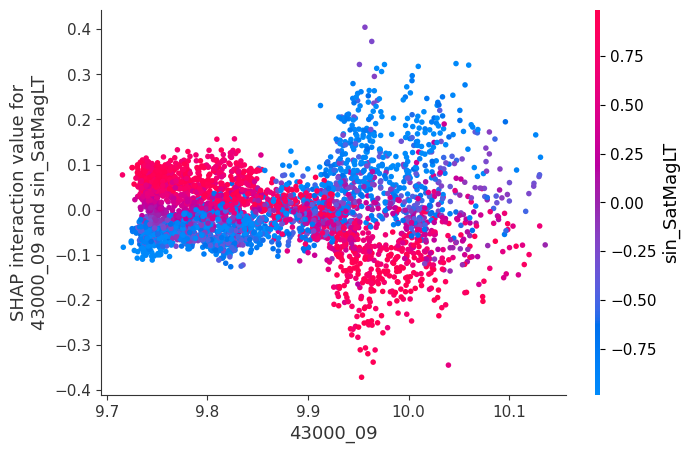}
            \caption{43nm vs. sin(MLT)}
            \label{fig:43sincross}
        \end{subfigure}
        \caption{Cross-interaction plots between FISM2 43nm input feature and selected other features. 
        X-axes go from low to high values of 43nm irradiance, and Y-axes go from low to high SHAP value. 
        Colors (if more than one) represent the magnitude of the cross-feature.
        For the cross-interaction between 43nm and 85nm (top right), note that the color (representing 85.55 band intensity) increases uniformly with the 43nm intensity.
        This indicates a high collinearity between the 43nm and 85.55nm input features in the model. 
        For the cross-interactions between 43nm and MLT, red colors represent night (dawn) and blue colors represent day (dusk) for cos(MLT) (sin(MLT)).
        \label{fig:cross43}}
\end{figure}

\begin{figure}
      \begin{subfigure}[b]{0.48\textwidth}
            \centering
            \includegraphics[width=\textwidth]{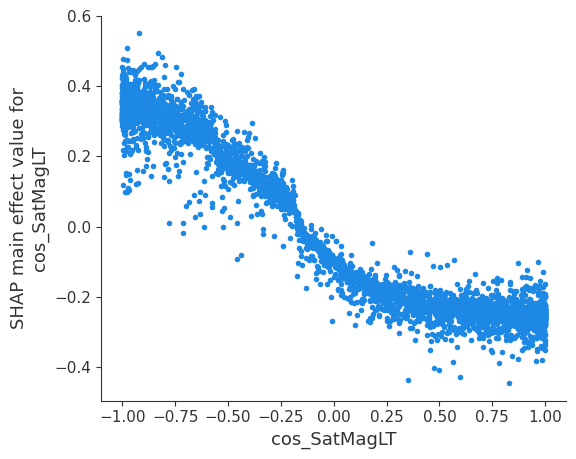}
            \caption{Main Feature for cos(MLT)}
            \label{fig:cosmain}
      \end{subfigure}
      \begin{subfigure}[b]{0.48\textwidth}
            \centering
            \includegraphics[width=\textwidth]{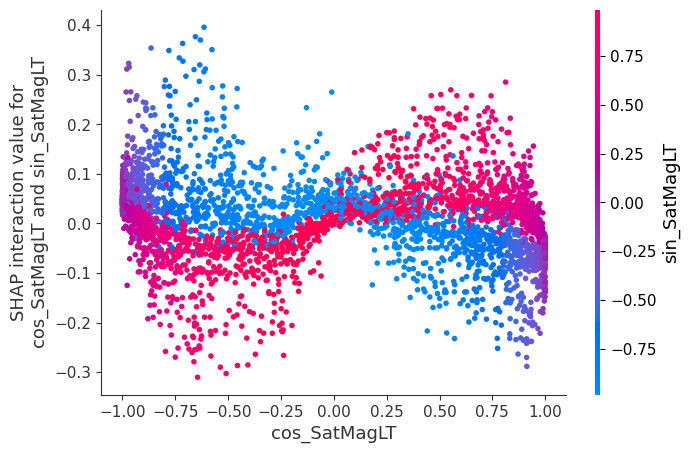}
            \caption{cos(MLT) vs. sin(MLT)}
            \label{fig:cossincross}
        \end{subfigure}
        \newline\centering
        \begin{subfigure}[b]{0.48\textwidth}
            \centering
            \includegraphics[width=\textwidth]{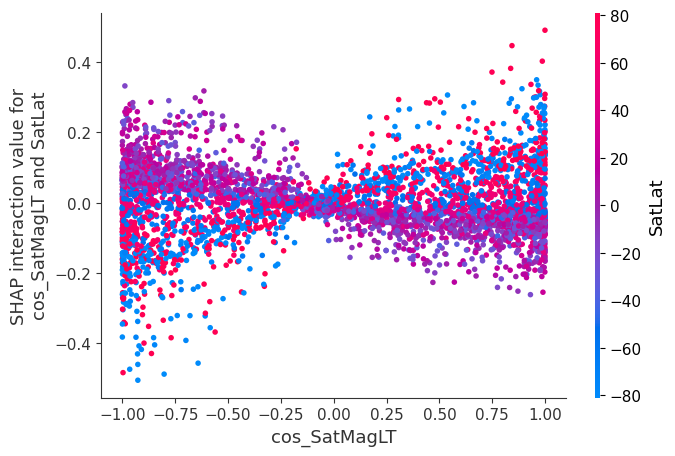}
            \caption{cos(MLT) vs. Satellite Latitude}
            \label{fig:cosLatcross}
        \end{subfigure}
        \caption{Main interaction for cos(MLT) and its cross-interaction with the sin(MLT) and SatLat features. 
        The X-axis goes left to right from local noon (cos(MLT) = -1) to local midnight (cos(MLT) = 1).
        The Y-axis goes from low to high SHAP value.
        Colors denote the magnitude of the second feature (blue is negative, red is positive). 
        \label{fig:crosscos}}
\end{figure}

The aggregate assessment shown in the beeswarm plot (Figure \ref{fig:beeswarmAll}) can be divided into main effects and interaction effects. 
The main effects represent the first-order relationships between feature magnitudes and their SHAP values. 
The interaction effects assess higher-order relationships between input features, which, when combined with the main effects, produce the final SHAP values.
These interactions are summarized in the interaction matrix (Figure \ref{fig:interaction}), which is created by summing the absolute values of the interaction SHAP values for every possible pair of input features. 
We observe that the FISM2 bands exhibit high interactions with one another, consistent with the input feature redundancies identified in the hierarchical analysis above (Figure \ref{fig:barAll}), especially between 43 nm and 85.55 nm. 

Selected examples for the 43nm band are shown in Figure \ref{fig:cross43}. 
The main effect (Figure \ref{fig:43main}) corroborates the results from the previous subsection: higher irradiance is associated with more positive SHAP values and higher atmospheric density.
Interestingly, there appear to be discrete ``jumps'' in SHAP values at certain irradiance thresholds.
The reason for this is unknown, and suggests a direction for future research.

The partial redundancy between FISM2 bands are evident in Figure \ref{fig:4385cross} where the 43nm irradiance is correlated with the 85.55nm irradiance, as indicated by the nearly monotonic relationship between the two features.
This indeed confirms the known issue with TreeSHAP values mentioned previously: features that have a minimal impact on the prediction may get a non-zero TreeSHAP estimate if highly linearly correlated with an important feature \cite{janzing2020, sundararajan2020}. Here, the 85.55nm feature appears to be very highly correlated with the 43nm feature, and as a result, obtains the third-highest SHAP importance overall.
However, from our hierarchical and cross-interaction analysis, we can understand that 85.55 nm overlaps significantly with 43 nm, and perhaps does not need to be included in the RANDM feature set.

Analyzing the cross-interactions between the 43nm band and the MLT features (Figures \ref{fig:43coscross} and \ref{fig:43sincross}) reveals both expected and unexpected relationships. 
As anticipated from the first-order effects of the 43 nm band and cos(MLT) (Figures \ref{fig:43main} and \ref{fig:cosmain}), higher densities are observed during daytime (positive cos(MLT)) and under high irradiance conditions.
Conversely, the model predicts lower densities at night for the same irradiance levels. 

However, at low irradiance levels, an opposite second-order effect emerges: the model shows a modest density enhancement during nighttime and a reduction during the daytime.
We speculate that this may be related to convective cells driven by $\vec{E}\times\vec{B}$ ion drifts, which are associated with dawn-dusk density asymmetries \cite{kwak2009, grocott2012, forster2017}.
When irradiance is low, ion production diminishes and the convective cells weaken, reducing the asymmetry between the day/dusk and night/dawn sectors.
This is reflected in the second-order effect observed here: at low irradiance, density increases in the night/dawn sectors and decreases in the day/dusk sectors, resulting in reduced density asymmetry between these regions.

We also see some interesting interactions between spatial locations (Figure \ref{fig:crosscos}).
Figure \ref{fig:cossincross} shows the day/night vs. dawn/dusk cross interactions. 
These mainly follow the established pattern: day and dusk have enhanced densities, night and dawn have lower densities.
Interestingly, there is a suggested higher-order effect between cos(MLT) and satellite latitude (Figure \ref{fig:cosLatcross}).
The equator (purple points in the figure) shows an expected relationship: day (night) is associated with higher (lower) densities. 
However, more polar latitudes (red/blue for north/south latitudes) show the opposite association!
In other words, polar latitudes on the nightside (dayside) are associated with higher (lower) densities compared to the equatorial latitudes.

\section{Local Assessment: Single Events}
\label{sec:local}

\begin{table}
\caption{Data associated with the individual ``Low", ``Medium", and ``High" events analyzed in Section \ref{sec:local}, as well as average values for the ``Quiet" and ``Storm" samples analyzed in Section \ref{sec:overview}.
I$\lambda$ represent logarithm of solar irradiance in the FISM2 band at the given wavelength at $\lambda$ nanometers, measured in log(photons/cm$^2$/sec).
Satellite Latitude is measured in degrees. ``Phase" is the storm phase, as defined by \protect\citeA{murphy2018,murphy2020}.
Density is the observed density at 400 kilometers, normalized to  $10^{-12}$ kg/$\mathrm{m}^3$\label{tab:evts}
}
\centering
\begin{tabular}{|c | c c c c c c c c c |c|}
\hline
      Event  & I1.3 & I43.00 & I85.55 & I94.40 & SYM-H & AE & Latitude & MLT & Phase & Density\\
\hline
      Low  &  7.16 & 9.82 &9.41 &9.19& -1.0 & 220.0 & 38.9 & 8.93 & Quiet & 0.944\\
      Medium & 7.39 & 9.93 & 9.47 & 9.23 & -44.0 & 169.0 & -14.0 & 6.20 & Recovery & 1.665   \\
      High  & 7.78 & 10.10 &9.57 &9.33& -25.0 & 112.0 & 43.1 & 9.64 & Recovery & 4.485 \\
      Avg. Quiet & 7.09 & 9.83 &9.42 & 9.19 & -5.0 & 104.0 & N/A& N/A & Quiet & 1.030\\
      Avg. Storm & 7.27 & 9.88 & 9.44 & 9.21 & -15.8 & 226.3 &N/A &N/A &Storm &1.535 \\
\hline
\end{tabular}
\end{table}

To demonstrate local SHAP analysis, we select three representative events (Table \ref{tab:evts}): 
a quiet-time event with low density (``Low") and two storm-time events, 
one with density predictions closer to the dataset's global mean (``Medium") and another with exceptionally high predicted density (``High").

\begin{figure}
      \begin{subfigure}[b]{0.48\textwidth}
            \centering
            \includegraphics[width=\textwidth]{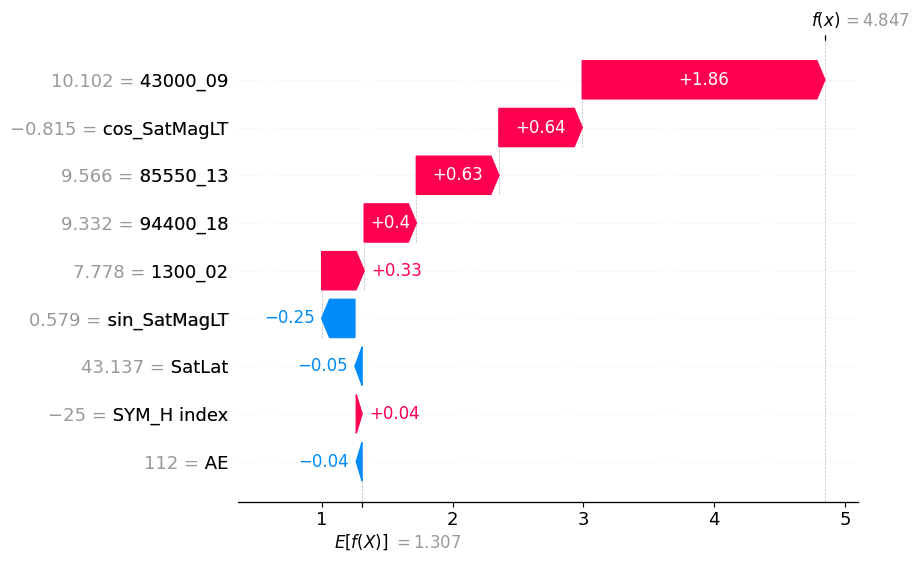}
            \caption{High Density Event}
            \label{fig:storm}
      \end{subfigure}
      \hfill
      \begin{subfigure}[b]{0.48\textwidth}
            \centering
            \includegraphics[width=\textwidth]{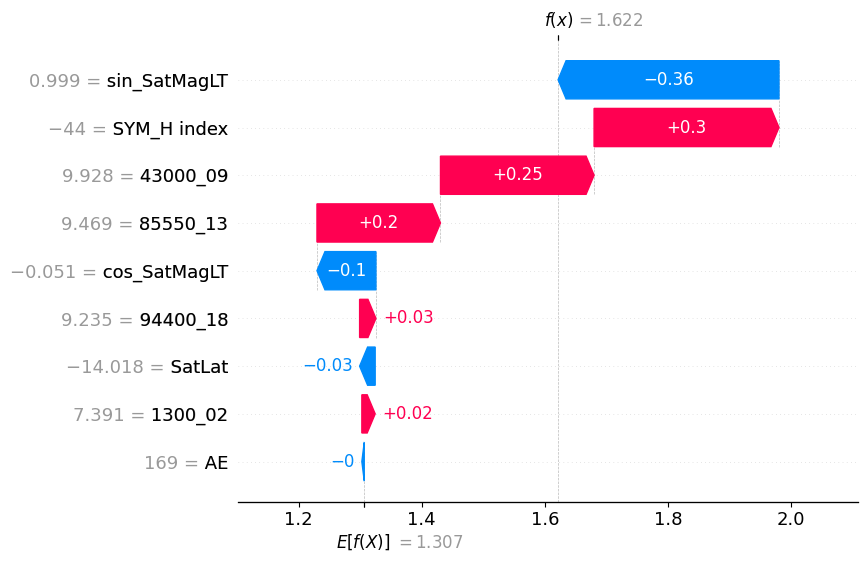}
            \caption{Medium Density Event}
            \label{fig:medium}
      \end{subfigure}
      \newline\centering
      \begin{subfigure}[b]{0.45\textwidth}
            \centering
            \includegraphics[width=\textwidth]{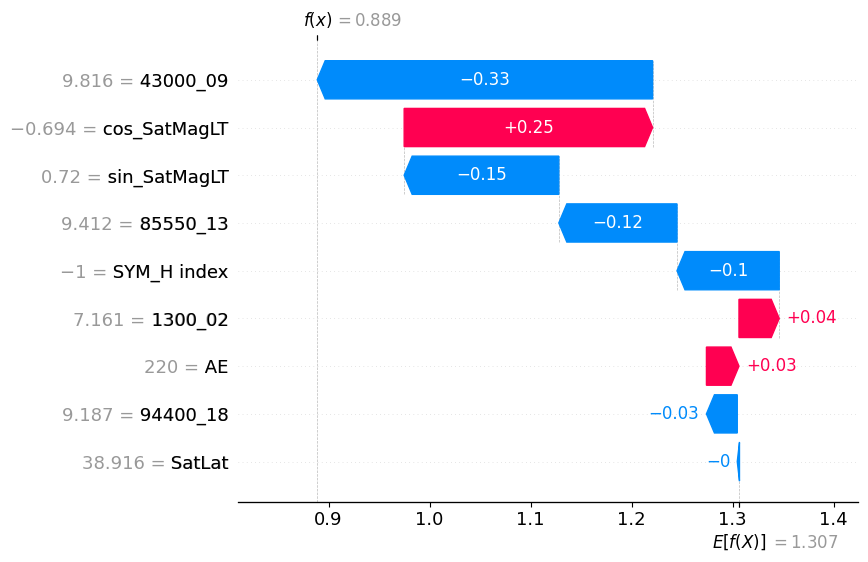}
            \caption{Low Density Event}
            \label{fig:low}
        \end{subfigure}
        \caption{Waterfall plots for High, Medium, and Low Density Events (details in text).\label{fig:waterfall}}
\end{figure}

Figure \ref{fig:waterfall} shows the SHAP values for each input feature for the three events.
The ``High" event occurred during a period of elevated solar irradiance, which the RF model associates with positive SHAP values for each of the FISM2 bands.
This combines with the satellite's position on the dayside (MLT = 9.64) to create a high density prediction.

Interestingly, the ``Medium" event took place during a time with a greater SYM-H index than the ``High" event, though with a lower background irradiance.
This is reflected in a higher SHAP value for SYM-H and a lower SHAP value for the FISM2 bands compared to the ``High'' event.
Additionally, the satellite's position at dawn (MLT $\approx$ 6) significantly decreases the final density prediction, as seen in the negative SHAP values for sin(MLT).
Finally, for the ``Low" event, although the observation point is on the dayside, the relatively low solar irradiance dominates the SHAP value assessment and lowers the final prediction.

\begin{figure}
      \includegraphics[width=0.9\textwidth]{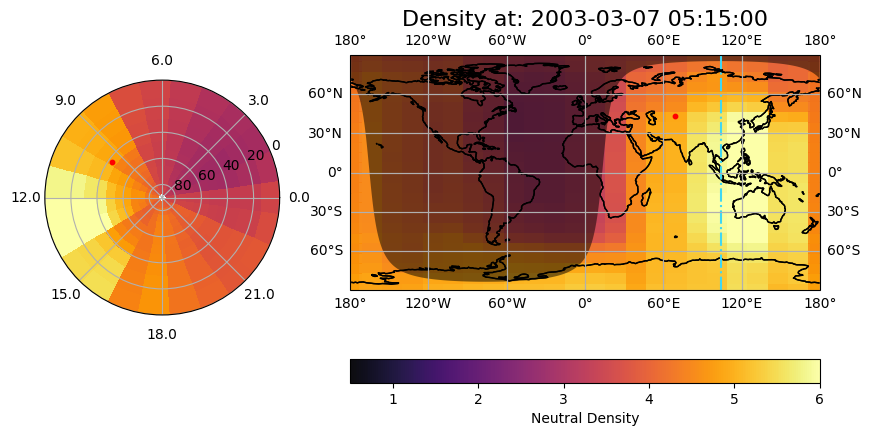}
      \caption{Predicted global density for the ``High" event, with density normalized to $10^{-12}~\mathrm{kg}/\mathrm{m}^3$
      The right map illustrates the predicted density overlaid on a geographic map in geographic coordinates, with the dot-dashed cyan line representing local noon (MLT = 12). This also shows the day-night terminator.
      The left panel depicts the predicted density in satellite latitude (SatLat) and magnetic local time (MLT) coordinates for the hemisphere in which the satellite is.
      The red dot indicates the satellite's location during this event.
      \label{fig:highGEO}}
\end{figure}

\subsection{Extension to Full-Earth Prediction}
\label{sec:extFull}
In the previous section, we demonstrated that local density predictions can be interpreted as combinations of input parameter importances.
We now extend this analysis to the full earth for individual events, allowing us to observe the interactions between input parameters and atmospheric location.
This will provide local views of the trends observed in the aggregate assessment of the full dataset (Section \ref{sec:macro}), including the cross-interactions (Section \ref{sec:crossInt}).

To achieve this, we define a grid of satellite latitude and MLT for a particular event, keeping all other event parameters fixed. 
Passing this grid into RANDM yields global density predictions associated with a single satellite measurement. 
Figure \ref{fig:highGEO} illustrates this extension to the entire globe for the ``High" event.
We observe a clear spatial correlation between daytime and increased density, with a bias toward dusk. 
This corroborates the results from Fig. \ref{fig:beeswarmAll}, where dayside and duskside MLTs are associated with positive SHAP values and generally higher densities. 

\begin{figure}
      \includegraphics[width=0.9\textwidth]{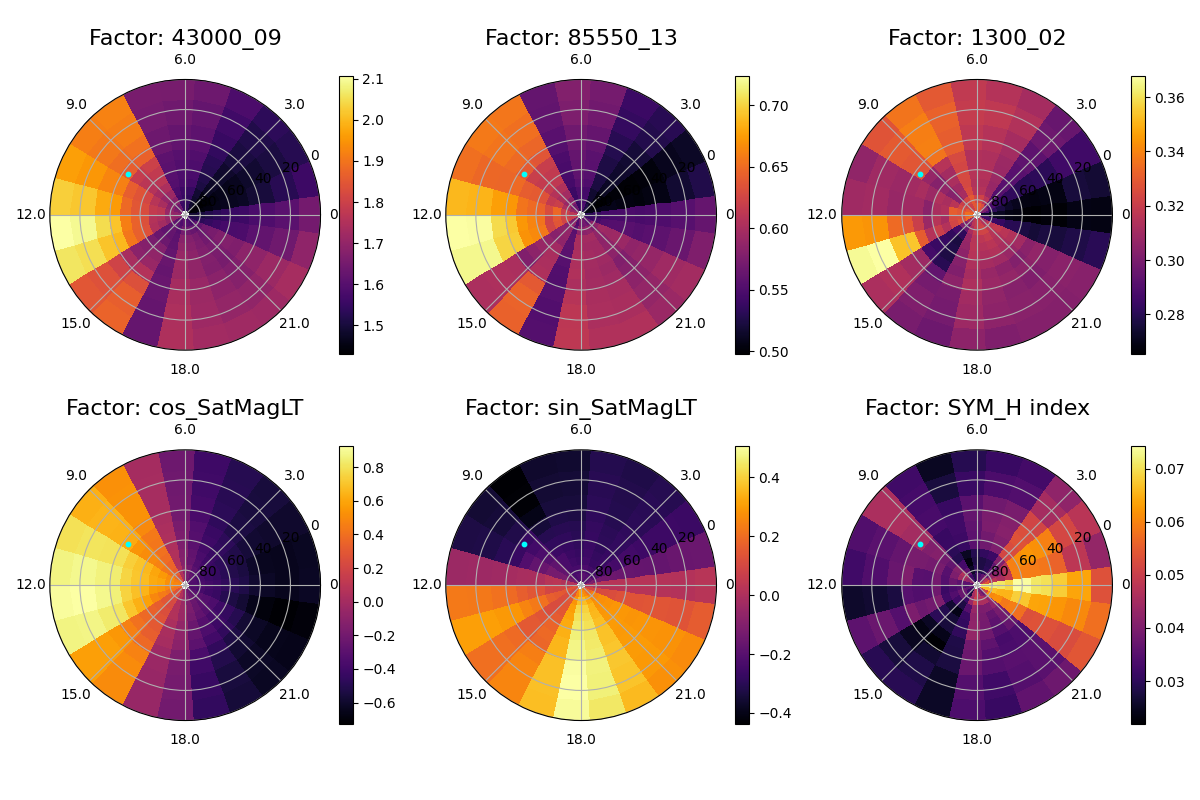}
      \caption{SHAP values for the given features across the North hemisphere for the ``High'' event.
      The cyan dot marks the satellite location for the original observation.
      \label{fig:globalHighSHAP}}
\end{figure}

\begin{figure}
      \includegraphics[width=0.9\textwidth]{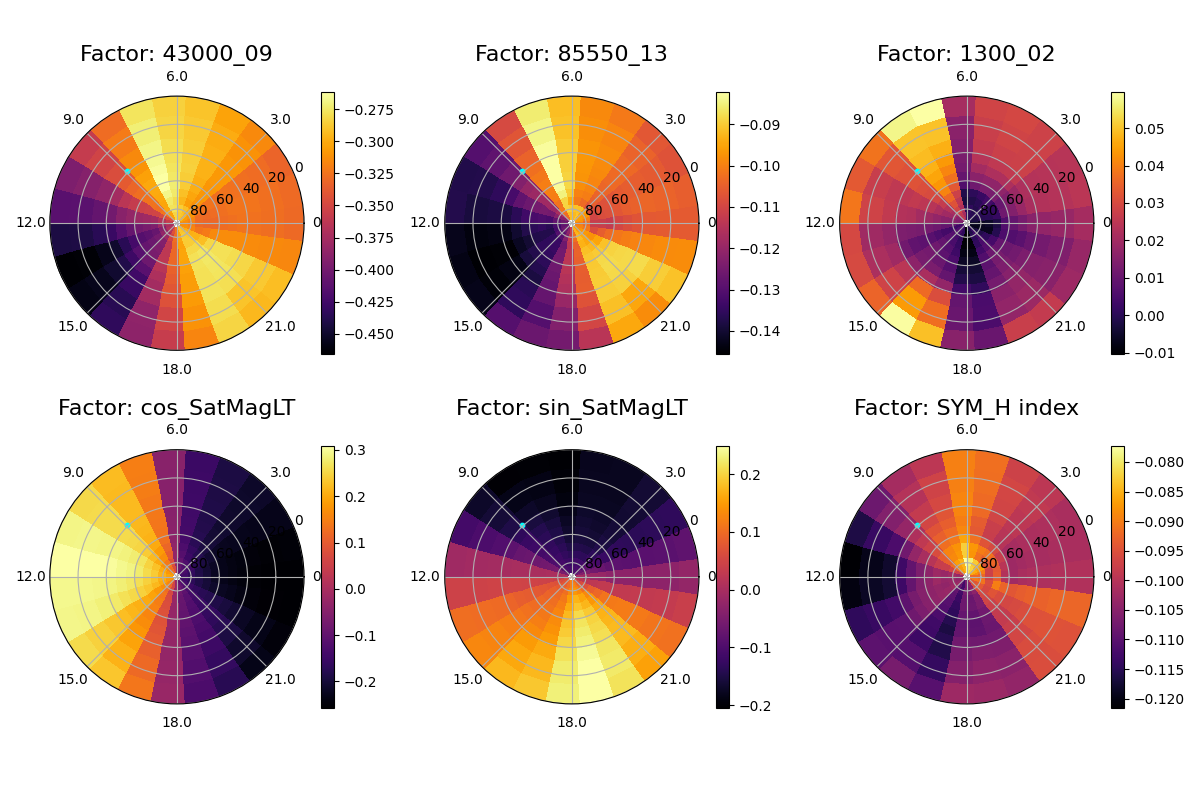}
      \caption{SHAP values for the given features across the North hemisphere for the ``Low'' event.
      The Cyan dot marks the satellite location for the original observation.
      \label{fig:globalLowSHAP}}
\end{figure}

Figures \ref{fig:globalHighSHAP} and \ref{fig:globalLowSHAP} show the SHAP distributions of the six most important features for the ``High'' and ``Low'' events. 
In the ``High'' event, the three irradiance SHAP plots indicate that the model predicts density enhancements across the entire hemisphere due to elevated irradiance levels. 
The higher SHAP values for irradiance are biased toward the dayside and dusk sectors for these events.
In contrast, for the ``Low'' event, the larger SHAP values for 43nm irradiance are biased toward the nightside.
This corroborates the interaction effects observed in Figures \ref{fig:43coscross} and \ref{fig:43sincross}, where a modest density enhancement during nighttime and a reduction during the day is observed at low irradiance.
Finally, all events consistently show a clear day/night and dawn/dusk split in the MLT SHAP values, reinforcing the influence of magnetic local time on density predictions.

\section{Conclusion}
\label{sec:conclude}
In this study, we applied SHAP (SHapley Additive exPlanations) analysis, utilizing the TreeSHAP algorithm, to a Random Forest model predicting thermospheric density based on solar-terrestrial parameters.
By interpreting the model's predictions through SHAP values, we gain insight into how different input features contribute to predicted atmospheric density variations.

Our analysis revealed that solar irradiance, particularly the 43nm spectral band, has a significant influence on density predictions. 
This finding underscores the role of specific wavelengths as key proxies for solar energy input into the thermosphere, corroborating the previous analysis in \citeA{murphy2025}.
All spectral bands used showed positive correlations with density; however, a hierarchical clustering analysis indicated redundancy among these input features.
This is unsurprising, as these are physically connected parameters.
This analysis suggests that future models may want to explore different combinations of proxies for solar energy input in order to minimize redundancies within this parameter space. 

Magnetic local time (MLT) components also demonstrated substantial impact on the model's predictions, with clear distinctions between the four quadrants (day, night, dawn, dusk).
The model associated higher densities with the dayside, aligning with the expected increase in atmospheric density due to direct solar irradiance. 
The observed dawn-dusk asymmetry aligns with previous studies \cite{kwak2009, grocott2012, forster2017} linking enhanced dusk densities (and reduced dawn densities) to ionospheric convection cells driven by $\vec{E}\times\vec{B}$ ion drifts.

When comparing different geomagnetic conditions, we found that the trends in feature importances remained consistent across storm-time and quiet-time periods, as well as between storm main-phase and recovery phases. 
The only significant difference was a higher SHAP value for the SYM-H index during storm conditions, which is consistent with the implied role of geomagnetic activity on atmospheric density.

In conclusion, the application of TreeSHAP to our Random Forest model has enhanced our understanding of both the model's internal workings and the underlying physical processes affecting thermospheric density. 
This methodology demonstrates the value of explainable machine learning techniques in space weather research, offering a pathway toward more interpretable and transparent models. 
Future work can build on this framework to incorporate additional parameters, explore other machine learning approaches, and apply similar analyses to different aspects of space weather phenomena.

%
%

\section*{Open Research Section}
The RANDM model \cite{murphy2025} is available under a MIT license at \url{https://github.com/kylermurphy/mltdm} and via Zenodo  \cite{murphy2025zenodo}: \url{https://zenodo.org/records/15091331}. Jupyter notebooks and supporting scripts for the SHAP analysis in this paper are hosted at \url{https://github.com/AIMFAHR/RANDM_SHAP_paper}.

\acknowledgments
Authors acknowledge funding from the NASA Heliophysics Internal Scientist Funding Model, specifically the AIMFAHR group. Software packages used in this research include: SHAP \cite{lundberg2020}, Fast Tree SHAP \cite{yang2021}, Numpy \cite{numpy2020}, Pandas \cite{pandas2010}, and MatplotLib \cite{matplotlib2007}.
The authors declare there are no conflicts of interest for this manuscript.

%
%

\bibliography{shap.bib}

\begin{thebibliography}{}

\bibitem [\protect \citeauthoryear {%
Berger%
, Holzinger%
, Sutton%
\BCBL {}\ \BBA {} Thayer%
}{%
Berger%
\ \protect \BOthers {.}}{%
{\protect \APACyear {2020}}%
}]{%
berger2020}
\APACinsertmetastar {%
berger2020}%
\begin{APACrefauthors}%
Berger, T\BPBI E.%
, Holzinger, M.%
, Sutton, E.%
\BCBL {}\ \BBA {} Thayer, J.%
\end{APACrefauthors}%
\unskip\
\newblock
\APACrefYearMonthDay{2020}{}{}.
\newblock
{\BBOQ}\APACrefatitle {Flying through uncertainty} {Flying through uncertainty}.{\BBCQ}
\newblock
\APACjournalVolNumPages{Space Weather}{18}{1}{e2019SW002373}.
\PrintBackRefs{\CurrentBib}

\bibitem [\protect \citeauthoryear {%
Cavus%
\ \protect \BOthers {.}}{%
Cavus%
\ \protect \BOthers {.}}{%
{\protect \APACyear {2025}}%
}]{%
cavus2025arxiv}
\APACinsertmetastar {%
cavus2025arxiv}%
\begin{APACrefauthors}%
Cavus, H.%
, Wang, J\BPBI T\BPBI L.%
, Singampalli, T\BPBI P\BPBI S.%
, Coban, G\BPBI C.%
, Zhang, H.%
, ur Raheem, A.%
\BCBL {}\ \BBA {} Wang, H.%
\end{APACrefauthors}%
\unskip\
\newblock
\APACrefYearMonthDay{2025}{}{}.
\newblock
\APACrefbtitle {An Interpretable Machine Learning Approach to Understanding the Relationships between Solar Flares and Source Active Regions.} {An interpretable machine learning approach to understanding the relationships between solar flares and source active regions.}
\newblock
\begin{APACrefURL} \url{https://arxiv.org/abs/2502.15066} \end{APACrefURL}
\PrintBackRefs{\CurrentBib}

\bibitem [\protect \citeauthoryear {%
Chamberlin%
\ \protect \BOthers {.}}{%
Chamberlin%
\ \protect \BOthers {.}}{%
{\protect \APACyear {2020}}%
}]{%
chamberlin2020}
\APACinsertmetastar {%
chamberlin2020}%
\begin{APACrefauthors}%
Chamberlin, P\BPBI C.%
, Eparvier, F\BPBI G.%
, Knoer, V.%
, Leise, H.%
, Pankratz, A.%
, Snow, M.%
\BDBL {}Woods, T\BPBI N.%
\end{APACrefauthors}%
\unskip\
\newblock
\APACrefYearMonthDay{2020}{}{}.
\newblock
{\BBOQ}\APACrefatitle {The flare irradiance spectral model-version 2 (FISM2)} {The flare irradiance spectral model-version 2 (fism2)}.{\BBCQ}
\newblock
\APACjournalVolNumPages{Space Weather}{18}{12}{e2020SW002588}.
\PrintBackRefs{\CurrentBib}

\bibitem [\protect \citeauthoryear {%
Coughlan%
\ \protect \BOthers {.}}{%
Coughlan%
\ \protect \BOthers {.}}{%
{\protect \APACyear {2023}}%
}]{%
coughlan2023}
\APACinsertmetastar {%
coughlan2023}%
\begin{APACrefauthors}%
Coughlan, M.%
, Keesee, A.%
, Pinto, V.%
, Mukundan, R.%
, Marchezi, J\BPBI P.%
, Johnson, J.%
\BDBL {}Hampton, D.%
\end{APACrefauthors}%
\unskip\
\newblock
\APACrefYearMonthDay{2023}{}{}.
\newblock
{\BBOQ}\APACrefatitle {Probabilistic forecasting of ground magnetic perturbation spikes at mid-latitude stations} {Probabilistic forecasting of ground magnetic perturbation spikes at mid-latitude stations}.{\BBCQ}
\newblock
\APACjournalVolNumPages{Space Weather}{21}{6}{e2023SW003446}.
\PrintBackRefs{\CurrentBib}

\bibitem [\protect \citeauthoryear {%
Emmert%
\ \protect \BOthers {.}}{%
Emmert%
\ \protect \BOthers {.}}{%
{\protect \APACyear {2021}}%
}]{%
emmert2021nrlmsis}
\APACinsertmetastar {%
emmert2021nrlmsis}%
\begin{APACrefauthors}%
Emmert, J\BPBI T.%
, Drob, D\BPBI P.%
, Picone, J\BPBI M.%
, Siskind, D\BPBI E.%
, Jones~Jr, M.%
, Mlynczak, M\BPBI G.%
\BDBL {}others%
\end{APACrefauthors}%
\unskip\
\newblock
\APACrefYearMonthDay{2021}{}{}.
\newblock
{\BBOQ}\APACrefatitle {NRLMSIS 2.0: A whole-atmosphere empirical model of temperature and neutral species densities} {Nrlmsis 2.0: A whole-atmosphere empirical model of temperature and neutral species densities}.{\BBCQ}
\newblock
\APACjournalVolNumPages{Earth and Space Science}{8}{3}{e2020EA001321}.
\PrintBackRefs{\CurrentBib}

\bibitem [\protect \citeauthoryear {%
F{\"o}rster%
, Doornbos%
\BCBL {}\ \BBA {} Haaland%
}{%
F{\"o}rster%
\ \protect \BOthers {.}}{%
{\protect \APACyear {2017}}%
}]{%
forster2017}
\APACinsertmetastar {%
forster2017}%
\begin{APACrefauthors}%
F{\"o}rster, M.%
, Doornbos, E.%
\BCBL {}\ \BBA {} Haaland, S.%
\end{APACrefauthors}%
\unskip\
\newblock
\APACrefYearMonthDay{2017}{}{}.
\newblock
{\BBOQ}\APACrefatitle {The Role of the Upper Atmosphere for Dawn-Dusk Differences in the Coupled Magnetosphere-Ionosphere-Thermosphere System} {The role of the upper atmosphere for dawn-dusk differences in the coupled magnetosphere-ionosphere-thermosphere system}.{\BBCQ}
\newblock
\APACjournalVolNumPages{Dawn-Dusk Asymmetries in Planetary Plasma Environments}{}{}{125--141}.
\PrintBackRefs{\CurrentBib}

\bibitem [\protect \citeauthoryear {%
Goldstein%
, Kapelner%
, Bleich%
\BCBL {}\ \BBA {} Pitkin%
}{%
Goldstein%
\ \protect \BOthers {.}}{%
{\protect \APACyear {2015}}%
}]{%
goldstein2015}
\APACinsertmetastar {%
goldstein2015}%
\begin{APACrefauthors}%
Goldstein, A.%
, Kapelner, A.%
, Bleich, J.%
\BCBL {}\ \BBA {} Pitkin, E.%
\end{APACrefauthors}%
\unskip\
\newblock
\APACrefYearMonthDay{2015}{}{}.
\newblock
{\BBOQ}\APACrefatitle {Peeking inside the black box: Visualizing statistical learning with plots of individual conditional expectation} {Peeking inside the black box: Visualizing statistical learning with plots of individual conditional expectation}.{\BBCQ}
\newblock
\APACjournalVolNumPages{journal of Computational and Graphical Statistics}{24}{1}{44--65}.
\PrintBackRefs{\CurrentBib}

\bibitem [\protect \citeauthoryear {%
Gonzalez%
\ \protect \BOthers {.}}{%
Gonzalez%
\ \protect \BOthers {.}}{%
{\protect \APACyear {1994}}%
}]{%
gonzalez1994}
\APACinsertmetastar {%
gonzalez1994}%
\begin{APACrefauthors}%
Gonzalez, W.%
, Joselyn, J\BHBI A.%
, Kamide, Y.%
, Kroehl, H\BPBI W.%
, Rostoker, G.%
, Tsurutani, B\BPBI T.%
\BCBL {}\ \BBA {} Vasyliunas, V.%
\end{APACrefauthors}%
\unskip\
\newblock
\APACrefYearMonthDay{1994}{}{}.
\newblock
{\BBOQ}\APACrefatitle {What is a geomagnetic storm?} {What is a geomagnetic storm?}{\BBCQ}
\newblock
\APACjournalVolNumPages{Journal of Geophysical Research: Space Physics}{99}{A4}{5771--5792}.
\PrintBackRefs{\CurrentBib}

\bibitem [\protect \citeauthoryear {%
Grocott%
, Milan%
, Imber%
, Lester%
\BCBL {}\ \BBA {} Yeoman%
}{%
Grocott%
\ \protect \BOthers {.}}{%
{\protect \APACyear {2012}}%
}]{%
grocott2012}
\APACinsertmetastar {%
grocott2012}%
\begin{APACrefauthors}%
Grocott, A.%
, Milan, S.%
, Imber, S.%
, Lester, M.%
\BCBL {}\ \BBA {} Yeoman, T.%
\end{APACrefauthors}%
\unskip\
\newblock
\APACrefYearMonthDay{2012}{}{}.
\newblock
{\BBOQ}\APACrefatitle {A quantitative deconstruction of the morphology of high-latitude ionospheric convection} {A quantitative deconstruction of the morphology of high-latitude ionospheric convection}.{\BBCQ}
\newblock
\APACjournalVolNumPages{Journal of Geophysical Research: Space Physics}{117}{A5}{}.
\PrintBackRefs{\CurrentBib}

\bibitem [\protect \citeauthoryear {%
Harris%
\ \protect \BOthers {.}}{%
Harris%
\ \protect \BOthers {.}}{%
{\protect \APACyear {2020}}%
}]{%
numpy2020}
\APACinsertmetastar {%
numpy2020}%
\begin{APACrefauthors}%
Harris, C\BPBI R.%
, Millman, K\BPBI J.%
, van~der Walt, S\BPBI J.%
, Gommers, R.%
, Virtanen, P.%
, Cournapeau, D.%
\BDBL {}Oliphant, T\BPBI E.%
\end{APACrefauthors}%
\unskip\
\newblock
\APACrefYearMonthDay{2020}{{\APACmonth{09}}}{}.
\newblock
{\BBOQ}\APACrefatitle {Array programming with {NumPy}} {Array programming with {NumPy}}.{\BBCQ}
\newblock
\APACjournalVolNumPages{Nature}{585}{7825}{357--362}.
\newblock
\begin{APACrefURL} \url{https://doi.org/10.1038/s41586-020-2649-2} \end{APACrefURL}
\newblock
\begin{APACrefDOI} \doi{10.1038/s41586-020-2649-2} \end{APACrefDOI}
\PrintBackRefs{\CurrentBib}

\bibitem [\protect \citeauthoryear {%
Hunter%
}{%
Hunter%
}{%
{\protect \APACyear {2007}}%
}]{%
matplotlib2007}
\APACinsertmetastar {%
matplotlib2007}%
\begin{APACrefauthors}%
Hunter, J\BPBI D.%
\end{APACrefauthors}%
\unskip\
\newblock
\APACrefYearMonthDay{2007}{}{}.
\newblock
{\BBOQ}\APACrefatitle {Matplotlib: A 2D graphics environment} {Matplotlib: A 2d graphics environment}.{\BBCQ}
\newblock
\APACjournalVolNumPages{Computing in Science \& Engineering}{9}{3}{90--95}.
\newblock
\begin{APACrefDOI} \doi{10.1109/MCSE.2007.55} \end{APACrefDOI}
\PrintBackRefs{\CurrentBib}

\bibitem [\protect \citeauthoryear {%
Hutchinson%
, Wright%
\BCBL {}\ \BBA {} Milan%
}{%
Hutchinson%
\ \protect \BOthers {.}}{%
{\protect \APACyear {2011}}%
}]{%
hutchinson2011}
\APACinsertmetastar {%
hutchinson2011}%
\begin{APACrefauthors}%
Hutchinson, J\BPBI A.%
, Wright, D.%
\BCBL {}\ \BBA {} Milan, S.%
\end{APACrefauthors}%
\unskip\
\newblock
\APACrefYearMonthDay{2011}{}{}.
\newblock
{\BBOQ}\APACrefatitle {Geomagnetic storms over the last solar cycle: A superposed epoch analysis} {Geomagnetic storms over the last solar cycle: A superposed epoch analysis}.{\BBCQ}
\newblock
\APACjournalVolNumPages{Journal of Geophysical Research: Space Physics}{116}{A9}{}.
\PrintBackRefs{\CurrentBib}

\bibitem [\protect \citeauthoryear {%
Janzing%
, Minorics%
\BCBL {}\ \BBA {} Bl{\"o}baum%
}{%
Janzing%
\ \protect \BOthers {.}}{%
{\protect \APACyear {2020}}%
}]{%
janzing2020}
\APACinsertmetastar {%
janzing2020}%
\begin{APACrefauthors}%
Janzing, D.%
, Minorics, L.%
\BCBL {}\ \BBA {} Bl{\"o}baum, P.%
\end{APACrefauthors}%
\unskip\
\newblock
\APACrefYearMonthDay{2020}{}{}.
\newblock
{\BBOQ}\APACrefatitle {Feature relevance quantification in explainable AI: A causal problem} {Feature relevance quantification in explainable ai: A causal problem}.{\BBCQ}
\newblock
\BIn{} \APACrefbtitle {International Conference on artificial intelligence and statistics} {International conference on artificial intelligence and statistics}\ (\BPGS\ 2907--2916).
\PrintBackRefs{\CurrentBib}

\bibitem [\protect \citeauthoryear {%
King%
\ \BBA {} Papitashvili%
}{%
King%
\ \BBA {} Papitashvili%
}{%
{\protect \APACyear {2005}}%
}]{%
king2005}
\APACinsertmetastar {%
king2005}%
\begin{APACrefauthors}%
King, J.%
\BCBT {}\ \BBA {} Papitashvili, N.%
\end{APACrefauthors}%
\unskip\
\newblock
\APACrefYearMonthDay{2005}{}{}.
\newblock
{\BBOQ}\APACrefatitle {Solar wind spatial scales in and comparisons of hourly Wind and ACE plasma and magnetic field data} {Solar wind spatial scales in and comparisons of hourly wind and ace plasma and magnetic field data}.{\BBCQ}
\newblock
\APACjournalVolNumPages{Journal of Geophysical Research: Space Physics}{110}{A2}{}.
\PrintBackRefs{\CurrentBib}

\bibitem [\protect \citeauthoryear {%
Kwak%
, Richmond%
, Deng%
, Forbes%
\BCBL {}\ \BBA {} Kim%
}{%
Kwak%
\ \protect \BOthers {.}}{%
{\protect \APACyear {2009}}%
}]{%
kwak2009}
\APACinsertmetastar {%
kwak2009}%
\begin{APACrefauthors}%
Kwak, Y\BHBI S.%
, Richmond, A\BPBI D.%
, Deng, Y.%
, Forbes, J\BPBI M.%
\BCBL {}\ \BBA {} Kim, K\BHBI H.%
\end{APACrefauthors}%
\unskip\
\newblock
\APACrefYearMonthDay{2009}{}{}.
\newblock
{\BBOQ}\APACrefatitle {Dependence of the high-latitude thermospheric densities on the interplanetary magnetic field} {Dependence of the high-latitude thermospheric densities on the interplanetary magnetic field}.{\BBCQ}
\newblock
\APACjournalVolNumPages{Journal of Geophysical Research: Space Physics}{114}{A5}{}.
\PrintBackRefs{\CurrentBib}

\bibitem [\protect \citeauthoryear {%
Lundberg%
\ \protect \BOthers {.}}{%
Lundberg%
\ \protect \BOthers {.}}{%
{\protect \APACyear {2020}}%
}]{%
lundberg2020}
\APACinsertmetastar {%
lundberg2020}%
\begin{APACrefauthors}%
Lundberg, S\BPBI M.%
, Erion, G.%
, Chen, H.%
, DeGrave, A.%
, Prutkin, J\BPBI M.%
, Nair, B.%
\BDBL {}Lee, S\BHBI I.%
\end{APACrefauthors}%
\unskip\
\newblock
\APACrefYearMonthDay{2020}{}{}.
\newblock
{\BBOQ}\APACrefatitle {From local explanations to global understanding with explainable AI for trees} {From local explanations to global understanding with explainable ai for trees}.{\BBCQ}
\newblock
\APACjournalVolNumPages{Nature machine intelligence}{2}{1}{56--67}.
\PrintBackRefs{\CurrentBib}

\bibitem [\protect \citeauthoryear {%
Lundberg%
\ \BBA {} Lee%
}{%
Lundberg%
\ \BBA {} Lee%
}{%
{\protect \APACyear {2017}}%
}]{%
lundberg2017}
\APACinsertmetastar {%
lundberg2017}%
\begin{APACrefauthors}%
Lundberg, S\BPBI M.%
\BCBT {}\ \BBA {} Lee, S\BHBI I.%
\end{APACrefauthors}%
\unskip\
\newblock
\APACrefYearMonthDay{2017}{}{}.
\newblock
{\BBOQ}\APACrefatitle {A unified approach to interpreting model predictions} {A unified approach to interpreting model predictions}.{\BBCQ}
\newblock
\APACjournalVolNumPages{Advances in neural information processing systems}{30}{}{}.
\PrintBackRefs{\CurrentBib}

\bibitem [\protect \citeauthoryear {%
Ma%
\ \protect \BOthers {.}}{%
Ma%
\ \protect \BOthers {.}}{%
{\protect \APACyear {2023}}%
}]{%
ma2023}
\APACinsertmetastar {%
ma2023}%
\begin{APACrefauthors}%
Ma, D.%
, Bortnik, J.%
, Chu, X.%
, Claudepierre, S\BPBI G.%
, Ma, Q.%
\BCBL {}\ \BBA {} Kellerman, A.%
\end{APACrefauthors}%
\unskip\
\newblock
\APACrefYearMonthDay{2023}{}{}.
\newblock
{\BBOQ}\APACrefatitle {Opening the black box of the radiation belt machine learning model} {Opening the black box of the radiation belt machine learning model}.{\BBCQ}
\newblock
\APACjournalVolNumPages{Space Weather}{21}{4}{e2022SW003339}.
\PrintBackRefs{\CurrentBib}

\bibitem [\protect \citeauthoryear {%
Ma%
, Bortnik%
, Ma%
, Hua%
\BCBL {}\ \BBA {} Chu%
}{%
Ma%
\ \protect \BOthers {.}}{%
{\protect \APACyear {2024}}%
}]{%
ma2024}
\APACinsertmetastar {%
ma2024}%
\begin{APACrefauthors}%
Ma, D.%
, Bortnik, J.%
, Ma, Q.%
, Hua, M.%
\BCBL {}\ \BBA {} Chu, X.%
\end{APACrefauthors}%
\unskip\
\newblock
\APACrefYearMonthDay{2024}{}{}.
\newblock
{\BBOQ}\APACrefatitle {Machine learning interpretability of outer radiation belt enhancement and depletion events} {Machine learning interpretability of outer radiation belt enhancement and depletion events}.{\BBCQ}
\newblock
\APACjournalVolNumPages{Geophysical Research Letters}{51}{1}{e2023GL106049}.
\PrintBackRefs{\CurrentBib}

\bibitem [\protect \citeauthoryear {%
Molnar%
}{%
Molnar%
}{%
{\protect \APACyear {2025}}%
}]{%
molnar2025}
\APACinsertmetastar {%
molnar2025}%
\begin{APACrefauthors}%
Molnar, C.%
\end{APACrefauthors}%
\unskip\
\newblock
\APACrefYear{2025}.
\newblock
\APACrefbtitle {Interpretable Machine Learning} {Interpretable machine learning}\ (\PrintOrdinal{3}\ \BEd).
\newblock
\begin{APACrefURL} \url{https://christophm.github.io/interpretable-ml-book} \end{APACrefURL}
\PrintBackRefs{\CurrentBib}

\bibitem [\protect \citeauthoryear {%
K.~Murphy%
, Halford%
, Bard%
\BCBL {}\ \BBA {} Rae%
}{%
K.~Murphy%
, Halford%
, Bard%
\BCBL {}\ \BBA {} Rae%
}{%
{\protect \APACyear {2025}}%
}]{%
murphy2025zenodo}
\APACinsertmetastar {%
murphy2025zenodo}%
\begin{APACrefauthors}%
Murphy, K.%
, Halford, A.%
, Bard, C.%
\BCBL {}\ \BBA {} Rae, J.%
\end{APACrefauthors}%
\unskip\
\newblock
\APACrefYearMonthDay{2025}{{\APACmonth{03}}}{}.
\newblock
\APACrefbtitle {kylermurphy/mltdm: Initial Release.} {kylermurphy/mltdm: Initial release.}
\newblock
\APACaddressPublisher{}{Zenodo}.
\newblock
\begin{APACrefURL} \url{https://doi.org/10.5281/zenodo.15091331} \end{APACrefURL}
\newblock
\begin{APACrefDOI} \doi{10.5281/zenodo.15091331} \end{APACrefDOI}
\PrintBackRefs{\CurrentBib}

\bibitem [\protect \citeauthoryear {%
K.~Murphy%
, Halford%
, Liu%
\BCBL {}\ \protect \BOthers {.}}{%
K.~Murphy%
, Halford%
, Liu%
\BCBL {}\ \protect \BOthers {.}}{%
{\protect \APACyear {2025}}%
}]{%
murphy2025}
\APACinsertmetastar {%
murphy2025}%
\begin{APACrefauthors}%
Murphy, K.%
, Halford, A\BPBI J.%
, Liu, V.%
, Klenzing, J.%
, Smith, J.%
, Garcia-Sage, K.%
\BDBL {}Rae, I\BPBI J.%
\end{APACrefauthors}%
\unskip\
\newblock
\APACrefYearMonthDay{2025}{}{}.
\newblock
{\BBOQ}\APACrefatitle {Understanding and modeling the dynamics of storm-time atmospheric neutral density using random forests} {Understanding and modeling the dynamics of storm-time atmospheric neutral density using random forests}.{\BBCQ}
\newblock
\APACjournalVolNumPages{Space Weather}{23}{1}{e2024SW003928}.
\PrintBackRefs{\CurrentBib}

\bibitem [\protect \citeauthoryear {%
K\BPBI R.~Murphy%
\ \protect \BOthers {.}}{%
K\BPBI R.~Murphy%
\ \protect \BOthers {.}}{%
{\protect \APACyear {2020}}%
}]{%
murphy2020}
\APACinsertmetastar {%
murphy2020}%
\begin{APACrefauthors}%
Murphy, K\BPBI R.%
, Mann, I\BPBI R.%
, Sibeck, D\BPBI G.%
, Rae, I\BPBI J.%
, Watt, C\BPBI E.%
, Ozeke, L\BPBI G.%
\BDBL {}Baker, D\BPBI N.%
\end{APACrefauthors}%
\unskip\
\newblock
\APACrefYearMonthDay{2020}{}{}.
\newblock
{\BBOQ}\APACrefatitle {A framework for understanding and quantifying the loss and acceleration of relativistic electrons in the outer radiation belt during geomagnetic storms} {A framework for understanding and quantifying the loss and acceleration of relativistic electrons in the outer radiation belt during geomagnetic storms}.{\BBCQ}
\newblock
\APACjournalVolNumPages{Space weather}{18}{5}{e2020SW002477}.
\PrintBackRefs{\CurrentBib}

\bibitem [\protect \citeauthoryear {%
K\BPBI R.~Murphy%
\ \protect \BOthers {.}}{%
K\BPBI R.~Murphy%
\ \protect \BOthers {.}}{%
{\protect \APACyear {2018}}%
}]{%
murphy2018}
\APACinsertmetastar {%
murphy2018}%
\begin{APACrefauthors}%
Murphy, K\BPBI R.%
, Watt, C.%
, Mann, I\BPBI R.%
, Jonathan~Rae, I.%
, Sibeck, D\BPBI G.%
, Boyd, A.%
\BDBL {}others%
\end{APACrefauthors}%
\unskip\
\newblock
\APACrefYearMonthDay{2018}{}{}.
\newblock
{\BBOQ}\APACrefatitle {The global statistical response of the outer radiation belt during geomagnetic storms} {The global statistical response of the outer radiation belt during geomagnetic storms}.{\BBCQ}
\newblock
\APACjournalVolNumPages{Geophysical Research Letters}{45}{9}{3783--3792}.
\PrintBackRefs{\CurrentBib}

\bibitem [\protect \citeauthoryear {%
Ribeiro%
, Singh%
\BCBL {}\ \BBA {} Guestrin%
}{%
Ribeiro%
\ \protect \BOthers {.}}{%
{\protect \APACyear {2016}}%
}]{%
ribeiro2016}
\APACinsertmetastar {%
ribeiro2016}%
\begin{APACrefauthors}%
Ribeiro, M\BPBI T.%
, Singh, S.%
\BCBL {}\ \BBA {} Guestrin, C.%
\end{APACrefauthors}%
\unskip\
\newblock
\APACrefYearMonthDay{2016}{}{}.
\newblock
{\BBOQ}\APACrefatitle {" Why should i trust you?" Explaining the predictions of any classifier} {" why should i trust you?" explaining the predictions of any classifier}.{\BBCQ}
\newblock
\BIn{} \APACrefbtitle {Proceedings of the 22nd ACM SIGKDD international conference on knowledge discovery and data mining} {Proceedings of the 22nd acm sigkdd international conference on knowledge discovery and data mining}\ (\BPGS\ 1135--1144).
\PrintBackRefs{\CurrentBib}

\bibitem [\protect \citeauthoryear {%
Shapley%
}{%
Shapley%
}{%
{\protect \APACyear {1953}}%
}]{%
shapley1953}
\APACinsertmetastar {%
shapley1953}%
\begin{APACrefauthors}%
Shapley, L\BPBI S.%
\end{APACrefauthors}%
\unskip\
\newblock
\APACrefYearMonthDay{1953}{}{}.
\newblock
{\BBOQ}\APACrefatitle {Stochastic games} {Stochastic games}.{\BBCQ}
\newblock
\APACjournalVolNumPages{Proceedings of the national academy of sciences}{39}{10}{1095--1100}.
\PrintBackRefs{\CurrentBib}

\bibitem [\protect \citeauthoryear {%
Solomon%
\ \BBA {} Qian%
}{%
Solomon%
\ \BBA {} Qian%
}{%
{\protect \APACyear {2005}}%
}]{%
solomon2005}
\APACinsertmetastar {%
solomon2005}%
\begin{APACrefauthors}%
Solomon, S\BPBI C.%
\BCBT {}\ \BBA {} Qian, L.%
\end{APACrefauthors}%
\unskip\
\newblock
\APACrefYearMonthDay{2005}{}{}.
\newblock
{\BBOQ}\APACrefatitle {Solar extreme-ultraviolet irradiance for general circulation models} {Solar extreme-ultraviolet irradiance for general circulation models}.{\BBCQ}
\newblock
\APACjournalVolNumPages{Journal of Geophysical Research: Space Physics}{110}{A10}{}.
\PrintBackRefs{\CurrentBib}

\bibitem [\protect \citeauthoryear {%
Sundararajan%
\ \BBA {} Najmi%
}{%
Sundararajan%
\ \BBA {} Najmi%
}{%
{\protect \APACyear {2020}}%
}]{%
sundararajan2020}
\APACinsertmetastar {%
sundararajan2020}%
\begin{APACrefauthors}%
Sundararajan, M.%
\BCBT {}\ \BBA {} Najmi, A.%
\end{APACrefauthors}%
\unskip\
\newblock
\APACrefYearMonthDay{2020}{}{}.
\newblock
{\BBOQ}\APACrefatitle {The many Shapley values for model explanation} {The many shapley values for model explanation}.{\BBCQ}
\newblock
\BIn{} \APACrefbtitle {International conference on machine learning} {International conference on machine learning}\ (\BPGS\ 9269--9278).
\PrintBackRefs{\CurrentBib}

\bibitem [\protect \citeauthoryear {%
E.~Sutton%
, Forbes%
\BCBL {}\ \BBA {} Nerem%
}{%
E.~Sutton%
\ \protect \BOthers {.}}{%
{\protect \APACyear {2005}}%
}]{%
sutton2005}
\APACinsertmetastar {%
sutton2005}%
\begin{APACrefauthors}%
Sutton, E.%
, Forbes, J.%
\BCBL {}\ \BBA {} Nerem, R.%
\end{APACrefauthors}%
\unskip\
\newblock
\APACrefYearMonthDay{2005}{}{}.
\newblock
{\BBOQ}\APACrefatitle {Global thermospheric neutral density and wind response to the severe 2003 geomagnetic storms from CHAMP accelerometer data} {Global thermospheric neutral density and wind response to the severe 2003 geomagnetic storms from champ accelerometer data}.{\BBCQ}
\newblock
\APACjournalVolNumPages{Journal of Geophysical Research: Space Physics}{110}{A9}{}.
\PrintBackRefs{\CurrentBib}

\bibitem [\protect \citeauthoryear {%
E\BPBI K.~Sutton%
}{%
E\BPBI K.~Sutton%
}{%
{\protect \APACyear {2009}}%
}]{%
sutton2009}
\APACinsertmetastar {%
sutton2009}%
\begin{APACrefauthors}%
Sutton, E\BPBI K.%
\end{APACrefauthors}%
\unskip\
\newblock
\APACrefYearMonthDay{2009}{}{}.
\newblock
{\BBOQ}\APACrefatitle {Normalized force coefficients for satellites with elongated shapes} {Normalized force coefficients for satellites with elongated shapes}.{\BBCQ}
\newblock
\APACjournalVolNumPages{Journal of Spacecraft and Rockets}{46}{1}{112--116}.
\PrintBackRefs{\CurrentBib}

\bibitem [\protect \citeauthoryear {%
Wahr%
, Swenson%
, Zlotnicki%
\BCBL {}\ \BBA {} Velicogna%
}{%
Wahr%
\ \protect \BOthers {.}}{%
{\protect \APACyear {2004}}%
}]{%
wahr2004}
\APACinsertmetastar {%
wahr2004}%
\begin{APACrefauthors}%
Wahr, J.%
, Swenson, S.%
, Zlotnicki, V.%
\BCBL {}\ \BBA {} Velicogna, I.%
\end{APACrefauthors}%
\unskip\
\newblock
\APACrefYearMonthDay{2004}{}{}.
\newblock
{\BBOQ}\APACrefatitle {Time-variable gravity from GRACE: First results} {Time-variable gravity from grace: First results}.{\BBCQ}
\newblock
\APACjournalVolNumPages{Geophysical Research Letters}{31}{11}{}.
\PrintBackRefs{\CurrentBib}

\bibitem [\protect \citeauthoryear {%
{W}es {M}c{K}inney%
}{%
{W}es {M}c{K}inney%
}{%
{\protect \APACyear {2010}}%
}]{%
pandas2010}
\APACinsertmetastar {%
pandas2010}%
\begin{APACrefauthors}%
{W}es {M}c{K}inney.%
\end{APACrefauthors}%
\unskip\
\newblock
\APACrefYearMonthDay{2010}{}{}.
\newblock
{\BBOQ}\APACrefatitle {{D}ata {S}tructures for {S}tatistical {C}omputing in {P}ython} {{D}ata {S}tructures for {S}tatistical {C}omputing in {P}ython}.{\BBCQ}
\newblock
\BIn{} {S}t\'efan van~der {W}alt\ \BBA {} {J}arrod {M}illman\ (\BEDS), \APACrefbtitle {{P}roceedings of the 9th {P}ython in {S}cience {C}onference} {{P}roceedings of the 9th {P}ython in {S}cience {C}onference}\ (\BPG~56 - 61).
\newblock
\begin{APACrefDOI} \doi{10.25080/Majora-92bf1922-00a} \end{APACrefDOI}
\PrintBackRefs{\CurrentBib}

\bibitem [\protect \citeauthoryear {%
Yang%
}{%
Yang%
}{%
{\protect \APACyear {2021}}%
}]{%
yang2021}
\APACinsertmetastar {%
yang2021}%
\begin{APACrefauthors}%
Yang, J.%
\end{APACrefauthors}%
\unskip\
\newblock
\APACrefYearMonthDay{2021}{}{}.
\newblock
{\BBOQ}\APACrefatitle {Fast TreeSHAP: Accelerating SHAP Value Computation for Trees} {Fast treeshap: Accelerating shap value computation for trees}.{\BBCQ}
\newblock
\APACjournalVolNumPages{arXiv preprint arXiv:2109.09847}{}{}{}.
\PrintBackRefs{\CurrentBib}

\end{thebibliography}

%
%
%
%
%

\end{document}